\documentclass[showpacs,amssymb,twocolumn,aps,nofootinbib]{revtex4}
\usepackage{amsmath}
\usepackage{amstext}
\usepackage{amsopn}
\usepackage{amsfonts}
\usepackage{amssymb}
\usepackage{bbm}
\usepackage{accents}
\usepackage{empheq}
\usepackage{graphicx}
\usepackage{epsf}
\usepackage{graphics}
\usepackage[latin1]{inputenc}

\begin{document}

\title{Large time behavior in an exactly soluble out of equilibrium model}

\author{Ashok K. Das$^{a,b}$ and J. Frenkel$^{c}$}
\affiliation{$^a$ Department of Physics and Astronomy, University of Rochester, Rochester, NY 14627-0171, USA}
\affiliation{$^b$ Saha Institute of Nuclear Physics, 1/AF Bidhannagar, Calcutta 700064, INDIA}
\affiliation{$^{c}$ Instituto de Física, Universidade de São Paulo, 05508-090, São Paulo, SP, BRAZIL}

\begin{abstract}
We study the behavior of out of equilibrium retarded, advanced and correlated Green's functions within the context of an exactly soluble (quenched) model. We show, to the lowest order, that even though the pinch singularities cancel, there is a residual linear dependence on the time interval (after the quench) in the correlated Green's function which may invalidate perturbation theory. We sum the perturbation series to all orders in this simple model and show explicitly that the complete Green's functions are well behaved even for large time intervals. The exact form of the correlated Green's function allows us to extract a manifestly positive distribution function, for large times after the quench, which has a memory of the frequency of the initial system before the quench.
\end{abstract}


\maketitle

\section{Introduction}

In the real time formalism \cite{das}, pinch singularities can arise in a perturbative calculation in quantum field theory at finite temperature if the integrand of an amplitude involves a product of the form (in momentum space)
\begin{equation}
\frac{1}{p^{2}-m^{2}+i\epsilon} \frac{1}{p^{2}-m^{2} -i\epsilon}\, f (p),\label{1}
\end{equation}
where $\epsilon$ is infinitesimal and $f(p)$ is assumed to be non vanishing on the mass shell. This can arise if the amplitude involves, for example, products of Green's functions of the type $G_{++} (p) G_{--}(p)$ or $G_{R} (p)G_{A}(p)$ where the notations will be explained shortly. These singularities pose a danger for the validity of perturbation theory and, therefore, have been the subject of active investigation for several years \cite{calzetta,boyanovsky,bedaque1,altherr,bedaque,greiner,millington}. Most of these discussions have been carried out in momentum space while a few have addressed the issue in the mixed space (where time coordinate is not Fourier transformed but space coordinates are). It is known from these studies that the pinch singularities are not present in equilibrium in a thermal field theory with doubled degrees of freedom \cite{landsman,weldon}. In out of equilibrium models, it has been argued, in general, and shown explicitly at one loop in a  simple (quenched) $\phi^{2n+1}$ model that the pinch singularities also cancel out for finite time intervals (after the quench). However, the pinch singularity leaves behind a residual dependence on the time interval which may still invalidate perturbation theory (for large time intervals) in out of equilibrium models. This can be seen qualitatively as follows. In the $\phi^{2n+1}$ theory in the mixed space, for example, the integration of the intermediate time coordinate interval can be  over the interval $t_{0} \leq z^{0} \leq x^{0}$ leading to a factor of the form
\begin{equation}
\frac{1 - e^{-\epsilon (x^{0}-t_{0})}}{\epsilon},\label{2}
\end{equation}
where $t_{0}$ is the quench time (when an interaction is introduced). It is clear from \eqref{2} that, as long as the interval $(x^{0}-t_{0})$ is finite, the pinch singularity cancels out and the term reduces to $(x^{0}-t_{0})$ in the limit $\epsilon\rightarrow 0^{+}$. However, for large time intervals, the singularity does not cancel which would reflect in the large time behavior of the model. Subsequently, it has been formally argued that in such theories if the imaginary part of the self-energy is finite, it leads to a finite separation between the poles in the propagator in the upper and the lower halves of the complex plane. This, not only leads to an absence of the pinch singularity, but also leads to a well behaved propagator for large times.

In this short note, we study how all these features explicitly manifest in a simple (quenched) out of equilibrium model which is exactly soluble. In particular, we show that no pinch singularity is present in a first order calculation of the physical Green's functions although there is a linear dependence on the time interval in the correlated Green's function as argued above. In this model, the propagator can be summed to all orders and the exact physical Green's functions show the following features. First, the retarded as well as the advanced Green's functions have well behaved large time behavior as would be expected. All the dependence on time intervals in the correlated Green's function sum nicely into trigonometric functions which are oscillatory and, therefore,  well behaved for large times. In our model, the self-energy does not have an imaginary part and shows that there are other mechanisms (besides having a finite imaginary part in the self-energy) which can make the exact propagator well behaved when summed to all orders. Since the Green's functions can be calculated exactly, it allows us to extract a form for the distribution, for large times, which is positive definite and which has a memory of the frequency of the initial system before the quench. 

Our paper is organized in the following manner. In section II, we recapitulate various Green's functions as well as various useful relations in the doubled theory in the real time (closed time path) formalism \cite{das}. In section III, we introduce our simple (soluble) model and point out its relevance to some other nontrivial models. We also present the first order corrections to various physical Green's functions in this section. In section IV, we derive the exact Green's functions for the theory (which would correspond to summing self-energy insertions to all orders in perturbation theory). We discuss various features which arise from this exact result.

\section{Green's functions}

Out of equilibrium systems at finite temperature are best studied in the closed time path formalism where the Feynman Green's functions have a $2\times 2$ matrix form \cite{das}
\begin{equation}
G = \begin{pmatrix}
G_{++} & G_{+-}\\
G_{-+}  & G_{--}
\end{pmatrix},\label{3}
\end{equation}
where the ``$\pm$" subscripts refer to the two real branches of the closed time path in the complex $t$-plane. The physical Green's functions which include the retarded ($G_{R}$), advanced ($G_{A}$) and correlated ($G_{c}$) Green's functions also have a $2\times 2$ matrix structure of the form
\begin{equation}
\widehat{G} = \begin{pmatrix}
0 & G_{A}\\
G_{R} & G_{c}
\end{pmatrix}.\label{4}
\end{equation}
The two sets of Green's functions in \eqref{3} and \eqref{4} are related through a unitary transformation $Q$ as
\begin{equation}
\widehat{G} = Q G Q^{\dagger},\quad Q = \frac{1}{\sqrt{2}}\begin{pmatrix}
1 & - 1\\
1 & 1
\end{pmatrix},\label{5}
\end{equation}
which leads to relations between the elements in the two matrices.

Similarly, we can also define the two point functions as
\begin{equation}
\Gamma_{2} = \begin{pmatrix}
\Gamma_{++} & - \Gamma_{+-}\\
-\Gamma_{-+} & \Gamma_{--}
\end{pmatrix},\quad \widehat{\Gamma}_{2} = \begin{pmatrix}
0 & \Gamma_{2 A}\\
\Gamma_{2 R} & \Gamma_{2 c}
\end{pmatrix},\label{6}
\end{equation}
which are also related by the same unitary matrix $Q$ as
\begin{equation}
\widehat{\Gamma}_{2} = Q\Gamma_{2} Q^{\dagger}.\label{7}
\end{equation}
We note here that the negative signs in the definition of the Feynman two point function in \eqref{6} reflects the fact that in the closed time path formalism, conventional time decreases along the direction of the path in the lower ($-$) branch. This also leads to the nonstandard inverse relation ($\sigma_{1}, \sigma_{3}$ correspond respectively  to the first and  the third of the Pauli matrices)
\begin{equation}
\Gamma_{2}\sigma_{3} G = \sigma_{3},\quad \widehat{\Gamma}_{2} \sigma_{1} \widehat{G} = \sigma_{1}. \label{8}
\end{equation}
These are operatorial relations which would explicitly correspond to integral equations. These relations can be written in the standard form by defining, for example, 
\begin{equation}
\overline{\Gamma}_{2} = \sigma_{3}\Gamma_{2}\sigma_{3},\label{9}
\end{equation}
which leads to $\overline{\Gamma}_{2} G = \mathbbm{1}$. For the physical two point function similarly, if we define 
\begin{equation}
\widehat{\overline{\Gamma}}_{2} = \sigma_{1}\widehat{\Gamma}_{2}\sigma_{1},\label{10}
\end{equation}
it would satisfy
\begin{equation}
\widehat{\overline{\Gamma}}_{2}\widehat{G} = \mathbbm{1}.\label{11}
\end{equation}

The analysis of the large time behavior of Green's functions is simpler in the physical basis and decomposing the two point function in \eqref{11} into a free part and the self-energy,
\begin{equation}
\widehat{\overline{\Gamma}}_{2} =  \left(\widehat{G}^{(0)}\right)^{-1} - \widehat{\overline{\Sigma}},\label{12}
\end{equation}
it follows from \eqref{11} that
\begin{equation}
\widehat{G} = (1 - \widehat{G}^{(0)} \widehat{\overline{\Sigma}})^{-1} \widehat{G}^{(0)}.\label{13}
\end{equation}
Taking various projections of this $2\times 2$ matrix we can determine the three complete physical Green's functions to have the forms
\begin{align}
G_{R} & =  (1 - G_{R}^{(0)}\Sigma_{R})^{-1} G_{R}^{(0)},\notag\\
G_{A} & = G_{A}^{(0)} (1 - \Sigma_{A} G_{A}^{(0)})^{-1},\label{14}\\
G_{c}  & = (1-G_{R}^{(0)}\Sigma_{R})^{-1} (G_{c}^{(0)} + G_{R}^{(0)}\Sigma_{c} G_{A}^{(0)}) (1-\Sigma_{A} G_{A}^{(0)})^{-1}.\notag
\end{align}
Since the retarded (advanced) Green's function involves only retarded (advanced) quantities, a little bit of analysis involving the boundary conditions shows that the integration over the intermediate time coordinates lie within a finite interval in this case and, therefore, should have a well behaved large time behavior. However, this is not quite obvious for the correlated Green's function.

\section{The model}

The model that we analyze is a simple exactly soluble model which has relevance to some more realistic models as well. We assume that for times $x^{0}\leq t_{0}$, where $t_{0}$ corresponds to a reference time, the system corresponds to a free scalar field theory of mass $m$ in equilibrium at temperature $T$. A mass correction  $\delta m^{2}$ is introduced at time $x^{0}=t_{0}$ and is present thereafter. Therefore, the Lagrangian density describing the system is given by
\begin{align}
{\cal L} & = \frac{1}{2} \partial_{\mu}\phi \partial^{\mu}\phi - \frac{m^{2}}{2} \phi^{2} - \theta (x^{0}-t_{0})\, \frac{\delta m^{2}}{2} \phi^{2}\notag\\
& = {\cal L}_{0} + {\cal L}_{I},\label{15}
\end{align}
where we treat the mass correction ($\delta m^{2}$) term as an interaction term and assume that
\begin{equation}
m^{2} + \delta m^{2} > 0.\label{16}
\end{equation}
Here the reference time $t_{0}$ defines the time of the quench which is normally taken to be $t_{0}=0$. However, we have left it arbitrary for purposes of generality. Even though the model is simple, our exact analysis for the Green's functions as well the corresponding conclusions  in this model hold even in more complicated models where the nontrivial interaction has the form $- \lambda \theta(x^{0}-t_{0}) \phi^{2n}, n\geq 2$ in the lowest loop (penguin) approximation.

From the form of the free theory, we can determine the free physical Green's functions which, in the mixed space, have the forms
\begin{align}
G_{R}^{(0)} (x^{0}-y^{0};\omega) & = - \frac{\theta(x^{0}-y^{0})}{\omega}\,\sin \omega (x^{0}-y^{0}),\notag\\
G_{A}^{(0)} (x^{0}-y^{0};\omega) & = \frac{\theta(y^{0}-x^{0})}{\omega}\,\sin \omega (x^{0}-y^{0}),\notag\\
G_{c}^{(0)} (x^{0}-y^{0};\omega) & = -\frac{i}{\omega}\left(1 + 2 n(\omega)\right) \cos \omega(x^{0}-y^{0}),\label{17}
\end{align}
where we have identified ($k=1$)
\begin{equation}
\omega = \sqrt{\mathbf{p}^{2}+m^{2}},\quad n(\omega) = \frac{1}{e^{\frac{\omega}{T}} - 1}.\label{18}
\end{equation}

Since the interaction consists of only a mass correction, the mass correction can be thought of as the exact self-energy in this model and we can correspondingly obtain
\begin{align}
& \Sigma_{R} (x^{0},y^{0};\omega) = \Sigma_{A}(x^{0},y^{0};\omega) = \theta (x^{0}-t_{0})\delta(x^{0}-y^{0})\delta m^{2},\notag\\
& \Sigma_{c} (x^{0}, y^{0};\omega) = 0.\label{19}
\end{align}
We note here that the self-energy, in this model, is completely real (there is no imaginary part). It can be checked that in the penguin approximation, this is also the form of the self-energy in models with a nontrivial interaction of the form $-\lambda \theta(x^{0}-t_{0}) \phi^{2n}, n\geq 2$.

We can now calculate first order corrections (with a single self-energy insertion) to the three tree level physical Green's functions. Using \eqref{14}, \eqref{17} and \eqref{18} we obtain
\begin{align}
G_{R}^{(1)} & = G_{R}^{(0)}\Sigma_{R}G_{R}^{(0)} = \frac{\theta(x^{0}-y^{0})\theta(y^{0}-t_{0})\delta m^{2}}{2\omega^{2}}\notag\\
& \times\left(\frac{\sin \omega(x^{0}-y^{0})}{\omega}  - (x^{0}-y^{0})\cos \omega (x^{0}-y^{0})\right),\notag\\
G_{A}^{(1)} & = G_{A}^{(0)}\Sigma_{A}G_{A}^{(0)} = -\frac{\theta(y^{0}-x^{0})\theta(x^{0}-t_{0})\delta m^{2}}{2\omega^{2}}\notag\\
& \times\left(\frac{\sin \omega(x^{0}-y^{0})}{\omega}  - (x^{0}-y^{0})\cos \omega (x^{0}-y^{0})\right),\notag\\
G_{c}^{(1)} & = G_{R}^{(0)}\Sigma_{R}G_{c}^{(0)} + G_{c}^{(0)}\Sigma_{A}G_{A}^{(0)}\notag\\
& = \frac{i\delta m^{2}}{2\omega^{2}}\left(1+2n(\omega)\right)\notag\\
&\times\Big[(\theta(x^{0}-t_{0})\left((x^{0}-t_{0}) \sin\omega (x^{0}-y^{0})\right.\notag\\
&\  + \frac{1}{2\omega}\left. (\cos \omega(x^{0}-y^{0}) - \cos\omega(x^{0}+y^{0}-2t_{0})\right)\notag\\
&\  - \theta(y^{0}-t_{0})\left((y^{0}-t_{0}) \sin\omega (x^{0}-y^{0})\right.\notag\\
&\  - \frac{1}{2\omega}\left. (\cos \omega(x^{0}-y^{0}) - \cos\omega(x^{0}+y^{0}-2t_{0})\right)\Big].\label{20}
\end{align}
As discussed above, it is clear that the retarded and the advanced Green's functions do not have any dependence on the fundamental time intervals $(x^{0}-t_{0})$and $(y^{0}-t_{0})$, but the correlated Green's function does and it is a residual reflection of the (cancelled) pinch singularity. For large values of the fundamental intervals, these contributions can grow and may pose a threat to perturbation theory. 

\section{Behavior of the exact Green's functions}

In this simple model, the Green's functions can be evaluated exactly with all order insertions of the self-energy using \eqref{14}. (This is also the case in models with the interaction of the form $-\lambda\theta(x^{0}-t_{0}) \phi^{2n}, n\geq 2$ in the penguin approximation.) In fact, since $\Sigma_{c}=0$ in our model, the calculation of the correlated Green's function is even simpler. 

First let us note that
\begin{align}
\lefteqn{1-G_{R}^{(0)}\Sigma_{R}}\notag\\
& = \delta (x^{0}-y^{0}) + \frac{\delta m^{2}\theta(x^{0}-y^{0})\theta(y^{0}-t_{0})}{\omega} \sin \omega(x^{0}-y^{0}).\label{21}
\end{align}
This allows us to determine the inverse to have the form
\begin{align}
\lefteqn{(1-G_{R}^{(0)}\Sigma_{R})^{-1}}\notag\\
& = \delta (x^{0}-y^{0}) - \frac{\delta m^{2}\theta(x^{0}-y^{0})\theta(y^{0}-t_{0})}{\Omega} \sin \Omega(x^{0}-y^{0}),\label{22}
\end{align}
where we have identified
\begin{equation}
\Omega^{2} = \omega^{2} + \delta m^{2}.\label{23}
\end{equation}
In a similar manner, we can also determine that
\begin{align}
\lefteqn{(1-\Sigma_{A}G_{A}^{(0)})^{-1}}\notag\\
& = \delta (x^{0}-y^{0}) + \frac{\delta m^{2}\theta(y^{0}-x^{0})\theta(x^{0}-t_{0})}{\Omega} \sin \Omega(x^{0}-y^{0}).\label{24}
\end{align}

Equations \eqref{22} and \eqref{24} (using \eqref{14}) allow us to determine the exact retarded and advanced Green's functions of the theory. For example, the retarded Green's function has the closed form
\begin{widetext}
\begin{align}
& G_{R} (x^{0},y^{0};\omega,\Omega) = \theta(t_{0}-x^{0})\theta(t_{0}-y^{0} G_{R}^{(0)} (x^{0}-y^{0};\omega) +\theta(x^{0}-t_{0})\theta(y^{0}-t_{0}) G_{R}^{(0)} (x^{0}-y^{0};\Omega)\notag\\
&\ +\frac{\theta(x^{0}-y^{0})\theta(x^{0}-t_{0})\theta(t_{0}-y^{0})}{2\omega}\left(\left(1-\frac{\omega}{\Omega}\right)\sin(\Omega(x^{0}-t_{0}) + \omega(y^{0}-t_{0})) - \left(1+\frac{\omega}{\Omega}\right)\sin (\Omega(x^{0}-t_{0})-\omega(y^{0}-t_{0}))\right),\label{25}
\end{align}
\end{widetext}
and the advanced Green's function can be obtained from \eqref{25} by letting $x^{0}\leftrightarrow y^{0}$. Both these Green's functions have the expected behavior for $x^{0},y^{0} < t_{0}$ as well as for $x^{0},y^{0} > t_{0}$  and are well behaved for large time intervals.

The complete correlated Green's function can also be calculated from \eqref{14} using \eqref{22} and \eqref{24} as well as the fact that $\Sigma_{c}=0$ and has the form
\begin{widetext}
\begin{align}
&G_{c} (x^{0},y^{0},\omega,\Omega) = - \frac{i}{\omega} (1+2 n(\omega)) \Big[\theta(t_{0}-x^{0})\theta(t_{0}-y^{0}) \cos \omega (x^{0}-y^{0})\notag\\
&\ +\frac{\theta(t_{0}-x^{0})\theta(y^{0}-t_{0})}{2}\left(\left(1+\frac{\omega}{\Omega}\right)\cos (\omega(x^{0}-t_{0})-\Omega(y^{0}-t_{0}))+\left(1-\frac{\omega}{\Omega}\right) \cos(\omega(x^{0}-t_{0}) + \Omega(y^{0}-t_{0})\right)\notag\\
&\ +\frac{\theta(x^{0}-t_{0})\theta(t_{0}-y^{0})}{2}\left(\left(1+\frac{\omega}{\Omega}\right)\cos (\Omega(x^{0}-t_{0})-\omega(y^{0}-t_{0}))+\left(1-\frac{\omega}{\Omega}\right) \cos(\Omega(x^{0}-t_{0}) + \omega(y^{0}-t_{0})\right)\notag\\
&\ + \frac{\theta(x^{0}-t_{0})\theta (y^{0}-t_{0})}{2}\left(\frac{(\Omega^{2}+\omega^{2})}{\Omega^{2}}\cos \Omega (x^{0}-y^{0}) + \frac{(\Omega^{2}-\omega^{2})}{\Omega^{2}} \cos \Omega(x^{0}+y^{0}-2t_{0})\right)\Big].\label{26}
\end{align}
\end{widetext}
The linear dependence on the intervals $(x^{0}-t_{0})$ and $(y^{0}-t_{0})$ pointed out at the first order of perturbation in \eqref{20} has nicely summed into oscillatory cosine functions. We note from \eqref{26} that, for $x^{0},y^{0}>t_{0}$, the correlated Green's function takes the form
\begin{align}
& G_{c} = - \frac{i}{2\omega} (1 + 2n (\omega))\left(\frac{(\Omega^{2}+\omega^{2})}{\Omega^{2}}\cos \Omega (x^{0}-y^{0})\right.\notag\\
& \qquad\left. + \frac{(\Omega^{2}-\omega^{2})}{\Omega^{2}} \cos \Omega(x^{0}+y^{0}-2t_{0})\right).\label{27}
\end{align} 
In particular, when the intervals $(x^{0}-t_{0})$ and $(y^{0}-t_{0})$ are large (long time after the quench), but $|x^{0}-y^{0}|$ is finite, the second term in \eqref{27} oscillates  rapidly around zero. In fact, from a careful calculation keeping the regularization parameter systematically, either in the mixed space \cite{oliver} or in momentum space, one finds that this term comes multiplied with a factor $e^{-\epsilon(x^{0} +y^{0}-2t_{0})}$. Here $\epsilon$ corresponds to the Feynman regularization parameter (see \eqref{1}) which is assumed to be taken to zero only at the end. Normally, such a factor can be set to unity in the limit $\epsilon\rightarrow 0^{+}$ when $\epsilon$ multiplies a finite interval. However, for large time intervals after the quench, this factor  provides a damping with a (damping) rate of the order $\epsilon$ and, therefore, in this case, we can neglect the second term in \eqref{27}. We note that the first term in \eqref{27} also comes multiplied with a multiplicative factor of the form $e^{-\epsilon |x^{0}-y^{0}|}$. However, since $|x^{0}-y^{0}|$ is finite, one can set this factor to unity in the limit $\epsilon\rightarrow 0^{+}$. Therefore, for large time intervals, we can  write \eqref{27}  as a correlated Green's function (see also \eqref{17})
\begin{equation}
G_{c} = -\frac{i}{\Omega} (1 + 2N(\omega,\Omega)) \cos \Omega (x^{0}-y^{0}),\label{28}
\end{equation}
with a positive definite distribution of the form
\begin{equation}
N(\omega, \Omega) = \frac{1}{2}\left[\frac{(\Omega^{2}+\omega^{2})}{2\omega\Omega}(1+2n(\omega)) -1\right],\label{29}
\end{equation}
which has a memory of the initial frequency $\omega$. 

We note that the time necessary for our system to come to thermal equilibrium is of the order of $\frac{1}{\epsilon}$, so that in the limit $\epsilon\rightarrow 0$, the system does not thermalize. The reason for this lies in the fact (which we have already pointed out) that in the class of models we are studying, the self-energy $\widehat{\Sigma}$ does not have any imaginary part which implies that there is no scattering (collisions). As a result, the system does not have any mechanism for  thermalization. In summary, our  analysis shows that it is not necessary to have a finite imaginary part of the self-energy for the Green's functions to have a good behavior for large times. They can be well behaved in an oscillatory manner when an imaginary part is not present although thermalization may not take place.
   
\bigskip

\noindent{\bf Acknowledgments}
\bigskip

 A. D. would like to thank the Departamento de F\'{i}sica Matem\'{a}tica in USP for hospitality where this work was done. This work was supported in part by USP and by CNPq (Brazil).


\begin{thebibliography}{10}

\bibitem{das} A. Das, {\em Finite Temperature Field Theory}, World Scientific, Singapore (1997).

\bibitem{calzetta} E. Calzetta and B. L. Hu, Phys. Rev. {\bf D37}, 2878 (1988).

\bibitem{boyanovsky} D, Boyanovsky, D. S. Lee and A. Singh, Phys. Rev. {\bf D48}, 800 (1993).

\bibitem{bedaque1} P. F. Bedaque and A. Das, Mod. Phys. Lett. {\bf A8}, 3151 (1993).

\bibitem{altherr} T. Altherr and D. Seibert, Phys. Lett. {\bf B333}, 149 (1994).

\bibitem{bedaque} P. F. Bedaque, Phys. Lett. {\bf B344}, 23 (1995).

\bibitem{greiner} C. Greiner and S. Leupold, Eur. Phys. Journ. {\bf C8}, 517 (1999).

\bibitem{millington} P. Millington and A. Pilaftsis, Phys. Rev. {\bf D88}, 085009 (2013).

\bibitem{landsman} N. P. Landsman and Ch. G. van Weert, Phys. Rep. {\bf 145}, 141 (1987).

\bibitem{weldon} H. A. Weldon, Phys. Rev. {\bf D45}, 352 (1992).

\bibitem{oliver} See, for example, eq. (83) in  F. T. Brandt, A. Das, O. Espinosa, J. Frenkel and S. Perez, Phys. Rev. {\bf D72}, 085006 (2005).




\end{thebibliography}
\end{document}